\documentclass[prl,amsmath,aps,superscriptaddress,twocolumn]{revtex4}
\usepackage{amsfonts}
\usepackage{graphicx,multirow}

\newcommand{\mX}{{\mathcal X}}
\newcommand{\mY}{{\mathcal Y}}

\newcommand{\mZ}{{\mathcal Z}}

\newcommand{\mW}{{\mathcal W}}

\newcommand{\mL}{{\mathcal L}}\newcommand{\mI}{{\mathcal I}}

\newcommand{\tr}{{\rm Tr}}

\newcommand{\bZ}{{\mathbb Z}}

\begin{document}

\title{Detecting the local indistinguishability of maximally entangled states}
\author{Sixia Yu}
\address{Centre for Quantum Technologies, National University of Singapore, 2 Science Drive 3, Singapore 117542}
\address{Hefei National Laboratory for Physical Sciences at
Microscale \& Department of Modern Physics,
 University of Science and Technology of China, Hefei, Anhui 230026, China}
\author{C.H. Oh}
\address{Centre for Quantum Technologies, National University of Singapore, 2 Science Drive 3, Singapore 117542}
\begin{abstract} 
By incorporating  the asymmetry of local protocols, i.e., some party has to start with a nontrivial measurement, into an operational method of detecting the local indistinguishability proposed by Horodecki {\it et al.} [Phys.Rev.Lett. 90 047902 (2003)], we derive a computable criterion to efficiently detect the local indistinguishability of maximally entangled states.  Locally indistinguishable sets of $d$ maximally entangled states in a $d\otimes d$ system are systematically constructed for all $d\ge 4$ as an application. Furthermore, by exploiting the fact that local protocols are necessarily separable, we explicitly construct  small sets of $k$ locally indistinguishable  maximally entangled states with the ratio $k/d$ approaching 3/4. In particular,  in a $d\otimes d$ system with even $d\ge 6$, there always exist $d-1$ maximally entangled states  that are locally indistinguishable by separable measurements.
\end{abstract}
\maketitle

{\it Introduction.--- } Not all properties of a composite system, typically those related to the entanglement,  can be accessed locally, i.e., by using only local operations and classical communication (LOCC). This is not surprising as entanglement and nonlocality are intimately related. It is thus striking when a complete set of orthogonal pure product states turns out to be locally indistinguishable \cite{ben99a}, i.e., the measurement of some observables with separable eigenstates cannot be implemented locally. This phenomenon of nonlocality without entanglement gives rise to the question as to what properties or global operations can or cannot be measured or implemented locally. The local identification of  orthogonal multipartite states, especially the maximally entangled states (MES), provides a perfect tool to explore this boundary.

Orthogonal multipartite states can always be distinguished by global measurements. Even restricted to the local protocols, two orthogonal pure states can always be exactly identified, regardless of the number of parties and whether or not the states are entangled \cite{wshv00}. However, locally indistinguishable sets of three or more  orthogonal states do exist, e.g., sets of pure product states\cite{ben99l,dms03,fs09} or  MESs \cite{gkrs04,nat05,ydy12,cos13,cr14,lwfz14} and even a mixture of them \cite{hssh03}. Due to the complex and elusive structures of LOCC protocols \cite{locc}, both the demonstration of the local distinguishability, for which one has to build  explicit local protocols, and the indistinguishability, for which one has to exclude all possible local protocols, are in general formidable tasks.

Despite our incomplete understanding, there do exist a few properties of  of LOCC protocols at our disposal to detect the local indistinguishability.  First, LOCC protocols are asymmetric \cite{gv01,wh02,nc06}, i.e., some party has to start with a nontrivial measurement. This seemly innocent property is highly nontrivial, leading to many important results and criteria \cite{cl03,yjc07,coh07}. Second, LOCC protocols cannot increase entanglement, which legitimates entanglement as a resource in various quantum informational tasks. A typical method is the state-identification induced entanglement transformation, the so-called HSSH method \cite{hssh03}, which develops from  a mixed state version in \cite{gkrss01,gkrsss02,yc03,gkrs04,fan04}. Third, LOCC protocols belong to some larger family of protocols, such as operations that are separable or having positive partial transpose (PPT), that are relatively well characterized \cite{cos13,cr14,ydy14,bcj14}.

In this Letter we shall at first combine the asymmetry and HSSH method to detect the local indistinguishability of MESs, called as  asymmetric HSSH method here, resulting in a computable criterion. As an application, we present the first complete construction of a locally indistinguishable set of $d$ maximally entangled states in a $d\otimes d$ system for all $d\ge4$. And then, by exploiting the fact that LOCC protocols are necessarily separable, we are also able to construct so far the smallest set of $k$ maximally entangled states in the case of $d$ being even with the ratio $k/d$ approaching 3/4. After introducing some necessary notations, we shall illustrate the asymmetry and HSSH methods by showing the local indistinguishability of a by far the smallest set of pure product states and a mixture of MESs and a  pure product state.

{\it Notations.--- } Here we shall consider the exact distinguishability of bipartite orthogonal pure states by finite LOCC protocols, i.e., by using local operations plus finite rounds of classical communication we would like to identify a state from a given set without any error. A set of mutually orthogonal pure states $\{|\psi_r\rangle\}$ is distinguished by a measurement $\{M_r\}$ if 
\begin{equation}
\langle\psi_{s}|M_r|\psi_{s}\rangle=\delta_{rs}.  
\end{equation}
If the measurement can be implemented by finite LOCC protocols or each $M_r$ is separable, then the set $\{|\psi_r\rangle\}$  is locally  distinguishable or distinguishable by separable measurement, respectively.

A bipartite system of two qudits, labeled with $A$ and $B$, is simply denoted by $d\otimes d$. For each  qudit, we denote by $\{|n\rangle\}$ with $n\in{\mathbb Z}_d:=\{0,1,\ldots,d-1\}$ its computational basis, by $I$ the identity operator, and by
\begin{equation}
X=\sum_{n\in {\mathbb Z}_d}|n+1\rangle\langle n|,\quad Z=\sum_{n\in {\mathbb Z}_d}\omega^n|n\rangle\langle n|,
\end{equation}
its bit and phase flip operators, respectively,  satisfying a Weyl-type commutation relation $ZX=\omega XZ$ with $\omega=e^{i2\pi/d}$. These operators are subscripted by the dimension they act on when necessary. In the case of qubit $d=2$ we shall denote by $\{\mI,\mX,\mY,\mZ\}$ the identity and three Pauli operators. In a  $d\otimes d$ system, let
\begin{equation}
|\Phi\rangle=\frac 1{\sqrt d}\sum_{n\in \bZ_d}|n\rangle\otimes |n\rangle
\end{equation}
denote the standard  maximally entangled state (MES). Each MES can be written as $|\psi_U\rangle=U\otimes I|\Phi\rangle$ for some unitary $U$ and we shall represent a set of MESs $\{|\psi_U\rangle\}$ also by the set $\{U\}$ of the corresponding unitaries. For convenience we shall denote by $\psi_U=|\psi_U\rangle\langle\psi_U|$ the density matrix of a pure state $|\psi_U\rangle$. For two orthogonal MESs represented by $U$ and $V$ it holds $\tr( UV^\dagger)=0$. The set of MESs $\{U_{st}=X^sZ^t\}$ is also referred to as MESs in canonical form or generalized Bell states.  

{\it Indistinguishability by asymmetry.--- } Local protocols for discrimination are asymmetric: some party has to start with a {\it nontrivial} and {\it non-disturbing} measurement, i.e., not all outcome $M_r$ is proportional to identity and after which the orthogonality relations are preserved, making further discrimination possible.  The method by asymmetry first appeared in \cite{gv01} and was elaborated in \cite{wh02,nc06} and further developed in \cite{yjc07,coh07}. 
Our first result is by far the smallest set of locally indistinguishable pure product states that demonstrates nonlocality without entanglement. 

{\bf Theorem 1} {\it In a $d\otimes d$ system with $d\ge 3$ the following  $2d-1$ orthogonal pure product states are locally indistinguishable:
\begin{equation}\label{pps}
\{|n\rangle\otimes |\delta_n\rangle\}_{n=1}^{d-1}\cup\{ |\delta_n\rangle\otimes|n_+\rangle\}_{n=1}^{d-1}\cup\{|\theta_0\rangle\otimes|\theta_0\rangle\},
\end{equation}
where $|\theta_0\rangle\propto\sum_j|j\rangle$ and $|\delta_n\rangle\propto|n\rangle-|0\rangle
$ and $n_+=n+1$ for $1\le n\le d-2$ while $n_+=1$ for $n=d-1$.}

{\bf Proof } We denote by $\{|\varphi_r\rangle\}$ those $2d-1$ pure product states and they are mutually orthogonal since $n_+\not=0,n$ and $|\delta_n\rangle$ is orthogonal to $|\theta_0\rangle$. Suppose that this set is locally distinguishable then someone, say, Alice, has to start with a nontrivial and nondisturbing measurement $\{M_A\}$, i.e., not all $M_A$ are proportional to identity and the post measurement states $\{\sqrt {M_A}\otimes I_B|\varphi_r\rangle\}$ should also be mutually orthogonal. As a result we have   $\langle n|M_A| n^\prime\rangle\langle\delta_n|\delta_{n^\prime}\rangle=0$ for all nonzero $n\not=n^\prime$, from which it follows that $\langle n|M_A| n^\prime\rangle=0$,   and $\langle n^\prime|M_A|\delta_n\rangle\langle\delta_{n^\prime}|n_+\rangle=0$, which in turn leads to $\langle n_+|M_A|0\rangle=0$ for all nonzero $n$ since $\langle\delta_{n^\prime}|n\rangle=0$  unless $n=n^\prime$. Thus all non diagonal elements of $M_A$ have to vanish. From the orthogonality $\langle\theta_0|M_A|\delta_k\rangle\langle\theta_0|n_+\rangle=0$  it follows that $\langle n|M_A|n\rangle=\langle 0|M_A|0\rangle$ for all $n\not=0$, i.e., $M_A$ is proportional to identify, meaning that Alice cannot start with a nontrivial measurement. For the same reason  Bob cannot start with a nontrivial measurement either. Therefore the set $\{|\varphi_r\rangle\}$ is locally indistinguishable.\hfill $\square$

After its discovery, nonlocality without entanglement is usually demonstrated by unextendible product basis (UPB) \cite{ben99l,dms03}, a set of mutually orthogonal pure product states that spans a subspace whose complementary subspace contains no pure product state.
Despite of being a natural generalization of a UPB in a $3\times 3$ system to higher dimensions, our set is extendible for all $d\ge 4$. The following $d-1$ mutually orthogonal pure  product states
$(|0\rangle+|n\rangle-2|(n_+)_{+}\rangle)\otimes |n_+\rangle$  for $n=1,2,\ldots d-1$ are orthogonal to all the pure product states in Eq.(\ref{pps}).
In the case of odd $d$ our set is of the same size as the minimal UPB while in the case of even $d$ our set is one state smaller than the smallest UPB, which   contains at least $2d$ states \cite{cj15}. Although the smallest number of locally indistinguishable  pure product states remains unknown, we conjecture that our sets are minimal, i.e., any set of no more than $2(d-1)$ product states is locally distinguishable, which is true for $d=3$ since  any 4 product states are shown to be locally distinguishable \cite{dms03}.

{\it HSSH method.--- }Local protocols cannot increase entanglement and local discrimination can be part of LOCC protocols, e.g., inducing a local  entanglement transfer between some specific pure states. The HSSH method \cite{hssh03} exploits this fact to detect the local indistinguishability when such a local state transformation is impossible, which is illustrated by our second result.

{\bf Theorem 2} {\it In a $d\otimes d$ system with $d\ge 3$ the following $k=[d/2]+1$ maximally entangled states
\begin{equation}
\{I,Z,Z^2,\ldots, Z^{[\frac d2]}\}
\end{equation}
together with one pure product state $|1\rangle\otimes |0\rangle$, are locally indistinguishable.}

{\bf Proof } We denote by $\Gamma=\{|\phi_r\rangle\}_{r=0}^{k}$ those $k+1$ states  given above, with the last one being the pure product state. By introducing two auxiliary qudits $C$ and $D$  
we build the so-called detector state, as the first step of HSSH method, i.e., a 4-qudit pure state, 
\begin{eqnarray}
|\Psi_\Gamma\rangle_{AC:BD}&=&\sum_{r=0}^{k-1}\sqrt{p_r}|\phi_r\rangle_{AB}\otimes |\psi_r\rangle_{CD}\cr&:=&dT_{AC}\otimes I_{BD}|\Phi\rangle_{AC:BD},
\end{eqnarray}
where $|\Phi\rangle_{AC:BD}=|\Phi\rangle_{AB}\otimes |\Phi\rangle_{CD}$, $|\psi_r\rangle=|\phi_r^*\rangle$ for $r=0,1,\ldots,k-1$ and $|\psi_k\rangle=|\psi_X\rangle=X\otimes I|\Phi\rangle$, and
\begin{equation}\label{t1}
T_{AC}=\frac 1d\sum_{r=0}^{k-1}\sqrt {p_r}Z^r\otimes Z^{-r}+\sqrt {\frac {p_{k}}d}|1\rangle\langle 0|\otimes X
\end{equation}
with  
$p_k=b$ and $p_r=a$ for $0\le r\le k-1$ with $ka+b=1$ and $b=d/(4k-d)<1$, recalling that $k>d/2$. Every local protocol successfully discriminating $\Gamma$, followed by a suitable local unitary transformation,  presents a local protocol of state transfer $\Psi_\Gamma\to\Phi$ in the $AC:BD$ cut. 

According to \cite{lp01, n99, vidal} the local state transfer $\Psi_\Gamma\to\Phi$ is possible if and only if $\lambda_1(\Psi_\Gamma)\le 1/d$   where  $\lambda_1(\Psi_\Gamma)$ denotes the largest Schmidt coefficient of $|\Psi_\Gamma\rangle_{AC:BD}$, i.e., the largest eigenvalue of
$M_{AC}= T_{AC}T_{AC}^\dagger $. However, in the 2-dimensional subspace spanned by $\{|0,0\rangle,|1,1\rangle\}_{AC}$ the matrix $M_{AC}$ has matrix elements
\begin{equation}
\tilde M_{AC}=\frac 1{d^2}\left(\begin{array}{cc} ak^2&k\sqrt{dab}\\ k\sqrt{dab}& ak^2+db\end{array}\right)
\end{equation}
from which it follows
$$\lambda_1(\Psi_\Gamma)\ge\lambda_1(\tilde M_{AC})
=\frac 1d+\frac{(2k-d)^2}{d^2(4k-d)}>\frac 1d,$$  
meaning that the local entanglement transfer $\Psi_\Gamma\to\Phi$, so that the local discrimination of $\Gamma$, is impossible. \hfill $\square$

We note that, first, if the pure product state is replaced by a MES, e.g., $Z^\dagger$, then the set becomes locally distinguishable, demonstrating a counterintuitive phenomenon of more nonlocality with less entanglement \cite{hssh03}. Second, as noted in \cite{ydy14}, the HSSH method alone cannot detect the local indistinguishability of $d$ MESs in a $d\otimes d$ system. Actually,
for any  set $\Gamma$  of $d$ MESs the following 4-partite pure state, which is the most general detector, 
\begin{equation}
|\Psi\rangle_{AC:BD}=\sum_{L\in\Gamma}\sqrt{p_L}|\psi_L\rangle_{AB}\otimes|\psi_{L^\prime}\rangle_{CD},
\end{equation}
where $\{|\psi_{L^\prime}\rangle\}$ are MESs in some auxiliary  systems $C$ and $D$ of dimension $d^\prime$, can always be transformed into $|\Phi\rangle_{CD}$ by LOCC. This is because the largest singular value of 
$T_{AC}=\sum_L\sqrt{p_L}L\otimes  L^\prime/\sqrt{dd^\prime}$
is at most $1/\sqrt {d^\prime}$, i.e., we always have $\lambda_1(\Psi)\le 1/d^\prime$, which ensures a local transfer of $\Psi$ into a MES \cite{lp01,n99}. 

{\it Asymmetric HSSH method.--- } We shall consider the local indistinguishability of a set of $k\le d$ MESs in  a $d\otimes d$ system, since any number $k>d$ of MESs cannot be locally distinguished \cite{nat05}, even by PPT measurements \cite{ydy12,cos13}. It turns out that any triplet of MESs in a $3\otimes 3$ system is locally distinguishable \cite{nat05} and in a $4\otimes 4$ there is a PPT indistinguishable quadruple of MESs \cite{ydy12}, which was generalized to the case of $d$ being a power of $2$ \cite{cos13}, for which small sets of  $k<d$ indistinguishable MESs were also constructed  \cite{cr14}.  An almost comprehensive construction of $d$ MESs in a $d\otimes d$ was provided  \cite{lwfz14} except for $d=5,11$ \cite{note}. Because of its relative small size i.e., $k\le d$, there are not enough orthogonality conditions to exclude a nontrivial measurement so that the asymmetry method alone does not work either.  However, a combination of those two methods above turns out to be extremely effective.

{\bf Theorem 3} {\it In a $d\otimes d$ system with $d\ge 4$ the following set of $d$ maximally entangled states
\begin{equation}
\Gamma_d=\{I,Z,Z^2,\ldots, Z^{d-3}, X^{[\frac d2]}, X^{\dagger[\frac d2]}Z^\dagger\}
\end{equation}
is locally indistinguishable.
}%

{\bf Proof } Suppose that the states are locally distinguishable and someone, say Alice, has to start with a nontrivial measurement $\{M\}$, i.e.,  there is at least one $M$ that is {\it not} proportional to the identity. After the $A$-measurement the set \{$|\psi_L\rangle\mid L\in\Gamma_d\}$ is transformed into  
\begin{equation}
\{|\phi_L\rangle=\sqrt{dM^\prime}\otimes I|\psi_L\rangle\mid L\in\Gamma_d\}, \; M^\prime=M/\tr M
\end{equation}
accordingly, which must also be mutually orthogonal in order to be distinguishable by further local protocols. 
By introducing two auxiliary qudits $C$ and $D$ we take the following 4-qudit pure state as the detector state
\begin{eqnarray}
|\Psi_{\Gamma_d}\rangle_{AC:BD}&:=&\frac1{\sqrt d}\sum_{L\in \Gamma_d}|\phi_L\rangle_{AB}\otimes |\psi_{L^*}\rangle_{CD}\cr&:=&dT_{AC}\otimes I_{BD}|\Phi\rangle_{AC:BD},
\end{eqnarray}
 where
$T_{AC}=\sum_{L\in \Gamma_d}(\sqrt {M^\prime} L)\otimes L^*/d$. The local entanglement transfer $\Psi_{\Gamma_d}\to\Phi$ is  possible if and only if the largest Schmidt coefficient $\lambda_1(\Psi_{\Gamma_d})\le 1/d$. 

Since  $\lambda_1(\Psi_{\Gamma_d})$  is given by the largest eigenvalue of $M_{AC}=T_{AC}T^\dagger_{AC}$,  we have a lower bound 
\begin{eqnarray}\label{mg}
\lambda_1(\Psi_{\Gamma_d})&\ge& \frac{\langle\varphi|M_{AC}|\varphi\rangle}{\langle\varphi|\varphi\rangle}=\frac1d+\frac{\tr( M_{\Gamma_d}-I)^2}{d^2} 
\end{eqnarray}
where  $|\varphi\rangle=\sqrt{M}\otimes I|\Phi\rangle$ and
$
M_{\Gamma_d}=\sum_{L\in \Gamma_d} {L^\dagger M^\prime L}.
$ 
If we can prove $M_{\Gamma_d}\not=I$ then we have $\lambda_1(\Psi_{\Gamma_d})>1/d$ so that  the local transfer, as well as the local discrimination, is impossible.
Here is exactly where the asymmetry enters into the play: we have only to show that for any nontrivial and non-disturbing $M$ it holds $M_{\Gamma_d}\not=I$, i.e.,  there exists $(s,t)\not=(0,0)$ with $s,t\in \bZ_d$ such that 
$0\not=\tr(M_\Gamma U_{st})={M_{st}\gamma_{st}}/\tr M$
where 
\begin{eqnarray}\label{gamma}
\gamma_{st}&:=&\frac1d\sum_{L\in \Gamma_d} \tr (U_{st}^\dagger L U_{st}L^\dagger )\cr&=&d\delta_{s0}-\omega^{-2s}-\omega^{-s}+\omega^{-[\frac d2]t}+\omega^{[\frac d2]t-s}
\end{eqnarray}
and $M_{st}=\tr(M U_{st})$, recalling that $U_{st}=X^sZ^t$. Since $M$ is nontrivial, i.e., there exists $(s,t)\not=(0,0)$ such that $M_{st}\not=0$,
it suffices to show that $\gamma_{st}\not=0$ or $\gamma_{st}=0$ infers $M_{st}=0$ for all $s,t\in\bZ_d$. In the case of odd $d$ we have $\gamma_{0t}\not=0$ since $d\ge 5$ and
$$\gamma_{st}=2\omega^{[\frac d2]s}\left(\cos\frac{2[\frac d2](s+t)}d\pi-\omega^{-s}\cos\frac{2[\frac d2]s}d\pi\right)\not=0$$
for all $t\in\bZ_d$ if $s\not=0$ since  $\omega^s$ is not real. As a result we have $\gamma_{st}\not=0$ for all $s,t\in\bZ_d$. 
In the case of even $d$  
$$\gamma_{st}=d\delta_{s0}+\big((-1)^t-\omega^{-s}\big)(1+\omega^{-s})$$ 
since $\omega^{d/2}=-1$.
If $d=4$ then $\gamma_{st}=0$ if and only if $(s,t)\in\{(0,1),(0,3),(2,t)\}$ and in these cases we have $M_{st}=0$ due to the fact that $\tr(ML^\prime L^\dagger)=0$ for different $L,L^\prime\in \Gamma_d$ as a result of  $\{|\phi_L\rangle\}$ being mutually orthogonal, e.g., $\tr(MZ^{\pm1})=0$. If  $d\ge 6$ then $\gamma_{st}=0$ if and only if $s=d/2$ with $t$ being arbitrary. From the orthogonality relationship of $\{|\phi_L\rangle\}$ it is straightforward to check that $\tr(MX^{d/2}Z^t)=0$ for all $t\in\bZ_d$, taking into account that $X^{d/2}$ is Hermitian. \hfill $\square$

Some remarks are in order. First, 
the proof above actually leads to a computable sufficient condition to detect the local indistinguishability of $d$ MESs in a $d\otimes d$ system.  We define a Weyl basis $\mW$ to be an  orthogonal unitary operator basis in which  $U_1U_2\propto U_2U_1$ for each pair of $U_{1,2}\in \mW$. The basis $\{U_{st}=X^sZ^t\}$ for a general qudit is one example. In the case of each subsystem bing a composite of multi qubits, all the multi qubit Pauli operators provide another example.

{\bf Lemma } {\it In a $d\otimes d$ system, a subset of a Weyl basis, i.e., $\mL\subset\mW$, defines a locally indistinguishable set of $d$ maximally entangled states  if  $|\mL|=d$
and
\begin{equation}\label{cd}
K(\mL)\subseteq\Delta(\mL)
\end{equation}
where $K(\mL)=\{U\in\mW|\gamma_{U}=0\}$ denotes the kernel set and $ 
\Delta(\mL):=\{L_1L_2^\dagger\mid L_1,L_2\in \mL\}
$ denotes the pairwise difference set with
\begin{equation}
\gamma_{U}=\frac1d\sum_{L\in\mL}\tr(L UL^\dagger U^\dagger ).
\end{equation}
}%

The proof of Lemma can proceed in exactly the same manner as the proof of Theorem 1 all the way to Eq(\ref{mg}), with $\Gamma_d$ replaced by $\mL$. The condition Eq.(\ref{cd}) ensures that a nontrivial $M$ leads to the existence of $U\in \mW$ with $U\not=I$ such that  $M_{U}\gamma_U\not=0$ which makes $M_\mL=\sum_{L\in\mL}L^\dagger ML/\tr M\not= I$ so that $\lambda_1(\Psi_\mL)>1/d$. Here, since $\mW$ is a basis, we have expansion $M=\sum_{U\in\mW}M_UU^\dagger/d$ with $M_U=\tr (MU)$.
For an example, in the case of even $d\ge 4$ the  set 
\begin{equation}
\Gamma_e=\{I,Z,Z^2,\ldots,Z^{d-2},X^{d/2}\}
\end{equation}
can be shown to be locally indistinguishable since  we have $K(\Gamma_e)=\{X^{dt/2}Z^t\mid t\not=0\}$ while $ \Delta(\Gamma_e)=\{X^{ds/2}Z^t\mid (s,t)\not=(0,0)\}$.
The indistinguishability of $\Gamma_e$ in the case of $d=4,6$  was conjectured \cite{gkrs04} and checked numerically \cite{cos13}. The local indistinguishability of five MESs $\{I,XZ,XZ^2,X^3Z,X^3Z^2\}$ in a $5\otimes 5$ system, which is verified numerically also in \cite{cos13}, can now be analytically proved since $K$ is an empty set. A  quadruple 
$\{I, XZ, XZ^3,X^2 Z^3\}$
is shown to be indistinguishable by one-way LOCC  protocols and conjectured to be locally indistinguishable \cite{bgk11}, which turns out to be true according to our criterion since we have $K=\{Z^2\}\subset\Delta$.

For the last example
we consider a $4\otimes 4$ system with each subsystem regarded as a composite system of two qubits. The identity and 15 Pauli operators of a 2-qubit system form a Weyl basis. The first example of LOCC indistinguishable set of $d$ MESs in a $d\otimes d$ system, i.e., a quadruple  
$\mL_4=\{\mI_1\mI_2,\mX_1\mX_2,\mY_1\mX_2,\mZ_1\mX_2\}$ of MESs  \cite{ydy12},
is a subset of this Weyl basis.
It is straightforward to check that $K({\mL_4})=\Delta({\mL_4})=\{\mX,\mY,\mZ\}\otimes \{\mI,\mX\}$ and the local indistinguishability of $\mL_4$ follows immediately from our Lemma.

Second,
unlike previous constructions of indistinguishable sets of MESs where the properties of PPT or separable measurements are employed, our construction deals with LOCC protocols directly. This makes it easier for a complete  construction of $d$ MES  in a $d\otimes d$ for all dimensions $d$ on the one hand and on the other hand we cannot exclude the possibility of being distinguished by some PPT or separable measurements, or even asymptotic LOCC protocols. 

{\it Small set of locally indistinguishable MESs.--- } Every LOCC protocol is separable so that it has positive partial transpose. This property has been used to construct small sets of $k<d$ locally indistinguishable MESs 
in a system $d\otimes d$ with $d$ being a power of 2 \cite{cr14}, which is significantly improved by our last result:

{\bf Theorem 4 } {\it In a $2d\otimes 2d$ system with $d\ge 2$ there exist $k_\sigma=2d-q+\sigma$ maximally entangled states that are indistinguishable by separable measurements, where   $q$ is the largest proper divisor of $d$ and $\sigma=1$ if $d$ is even and $0$ if $d$ is odd.}

{\bf Proof } For any $d\ge2$ with $q$ being its largest proper divisor, i.e., $q\not=d$ being the largest integer that divides $d$, there is a prime $p\ge 2$ such that $d=pq$ and $q\ge p$ if $q\not=1$. Each subsystem can be  regarded as a composite system of a qubit and two qudits with $p$ and $q$ levels. We claim that the following set of $k_\sigma=2d-q+\sigma$ MESs is indistinguishable by separable measurements:
\begin{equation}
\Xi_{2d}=\{Z^n_q\otimes L_V\mid n\in \bZ_q,V\in\mL_{p}^\sigma\},
\end{equation} 
where $L_V=|0\rangle\langle 1|\otimes V-|1\rangle\langle 0|\otimes V^T$ and
\begin{equation}
\mL_{p}^\sigma=\{Z_p^a\}_{a=0}^{p-2+\sigma}\cup\{X_pZ^a_p\}_{a\in\bZ_p},
\end{equation}
because of the following contradiction
\begin{eqnarray}\label{cds}
0&=&k_\sigma-\sum_{U\in\Xi_{2d}}\tr(M_U\psi_U)\cr
&=&k_\sigma-{\tr H_{2d}}+\sum_{U\in\Xi_{2d}}\tr M_U(H_{2d}-\psi_U)\cr
 &\ge&\sum_{U=Z_q^n\otimes L_V\in\Xi_{2d}}\tr M_U(P_q\otimes H_{L_V})+\sigma>0,
\end{eqnarray}
where we have denoted  $H_{2d}=P_q\otimes A_{2p}$ and $H_{L_V}=A_{2p}-\psi_{L_V}$ with
\begin{equation}
P_q=\sum_{n\in\bZ_q}|n,n\rangle\langle n,n|,\quad A_{2p}=\frac{I_{2p}\otimes I_{2p}-V_{2p}}{2p}
\end{equation}
and $V_{2p}$ being the swap operator on the $2p\otimes 2p$ system.

Suppose that the set $\Xi_{2d}$ can be distinguished by some separable measurement
$\{M_U\}_{U\in\Xi_{2d}}$, i.e., $\tr (M_U\psi_{U^\prime})=\delta_{UU^\prime}$ for arbitrary $U,U^\prime\in\Xi_{2d}$, from which the first equality in Eq.(\ref{cds}) follows immediately by noting $|\Xi_{2d}|=k_\sigma$.
 The second equality is due to the completeness of the measurement.
 The first inequality holds because $\tr H_{2d}=2d-q$ and $P_q=\sum_{n\in\bZ_q}\psi_{Z^n_q}$ so that
$\psi_U=\psi_{Z^n_q}\otimes\psi_{L_V}\le P_q\otimes \psi_{L_V}$. The last inequality holds because we have, firstly,
\begin{equation}\label{pme}
\tr M(P_q\otimes H_{L_V})\ge0,\;  \forall\ L_V\in\mL_{p}^\sigma,
\end{equation}
for any separable $M\ge0$ and, which immediately proves the theorem in the case of even $d$, and secondly,  
\begin{equation}\label{pmo}
\tr M_{I_q\otimes L_I}(P_q\otimes H_{L_I})>0
\end{equation} in the case of odd $d$, both of which will be proved in Appendix.  Actually $H_{L_V}$, as well as $P_d\otimes H_{L_V}$, defines an entanglement detecting positive map \cite{breuer06,hall06} which has been  used in \cite{bcj14} to detect indistinguishability by separable measurements in the case of $p=2$. \hfill $\square$

The minimal size of a locally indistinguishable sets of MESs inferred form Theorem 3 and 4 is summarized in Table I. Notably, in a $4m\otimes 4m$  system with $m\ge 1$ there is a set of  $k=3m+1$ locally indistinguishable MESs. Specially, in a $8m\otimes 8m$ system with $m\ge 1$ there exist $6m+1$,  instead of $7m+1$ in \cite{cr14}, locally indistinguishable MESs.  In the limit of large $d$ the ratio  $k/d$ approaches  $3/4$. As another consequence, in the case of even $d\ge 6$ we have $q-\sigma\ge 1$ so that  there always exists a set of $d-1$ MESs that is locally indistinguishable. 

\begin{table}
\begin{tabular}{ccl}
\hline\hline
$d$ & & $k_{min}$ \\
\hline
4 or $p\ge 5$ ( prime )&& $d$\\
$2p$ ($p\ge 3$ prime) && $d-1$\\
$4m$ $(m\ge 2)$ & & $\frac34d+1$\\
$6m$ $(m\ge 1$ odd) & &$\frac56d$\\
$2pq$ ($p\ge 5$ prime, $q\ge p$ odd) && $\frac{2p-1}{2p}d$\\
\hline\hline
\end{tabular}
\caption{The smallest size $k_{min}$ of the locally indistinguishable sets of MESs in all possible local dimension $d$ as constructed  form Theorem 3 and 4.}
\end{table}

{\it Conclusions and discussions.--- } We have exploited various properties of LOCC protocols to detect the exact local distinguishability of maximally entangled states. A computable criterion is derived by which many previously conjectured or only numerically checked indistinguishable sets of MESs are confirmed.  A complete construction of $d$ MESs in a $d\otimes d$ system  is provided for all $d\ge 4$ for the first time as well as small sets of locally indistinguishable MESs comparing to the local dimension.  Our method may also help in investigating the distinguishability by asymptotic LOCC protocols or unambiguous discrimination. 
Detection of the indistinguishability is only the first step, showing that there is nonlocality somewhere. The next step is to quantify the necessary nonlocal resource, such as entanglement, to complete the task of local discrimination. 

{\it Acknowledgement.--- }  This work is funded by the Singapore Ministry of Education (partly through the Academic Research Fund Tier 3 MOE2012-T3-1-009) and the National Research
Foundation, Singapore (Grant No. WBS: R-710-000-008-271).

\vskip 0.5cm

{\it Appendix: Proof of Eq.(\ref{pme}) and Eq.(\ref{pmo}).--- } Since $L_V$ is antisymmetric, i.e., $L^T_V=-L_V$ for all $V\in\mL_p^\sigma$, and $|y^*\rangle\langle y|$  is symmetric for an arbitrary state $|y\rangle$ in the $2 p$ system, we have $\langle y|L_V|y^*\rangle=0$, from which it follows 
\begin{equation}\label{h1}
2p\tr(x\otimes y)H_{L_V}=1-|\langle x|y\rangle|^2-|\langle x|L_V|y^*\rangle|^2\ge0
\end{equation}
 for arbitrary two normalized pure states $|x\rangle$ and $|y\rangle$ in the $2p$ system.
 As a result,  for an arbitrary pure state $|z\rangle_A=\sum_n|n\rangle\otimes|x_n\rangle$ and $|w\rangle_B=\sum_n|n\rangle\otimes|y_n\rangle$ on each subsystem $A$ and $B$ it holds 
\begin{equation}\label{h2}
\tr(z_A\otimes w_B)(P_q\otimes H_{L_V})=\sum_{n\in\bZ_q}
\tr(x_n\otimes y_n)H_{L_V}\ge 0
\end{equation}
since $\tr_{q}(z_A\otimes w_B)P_q=\sum_nx_n\otimes y_n$, with the trace taken over the $q$ systems from both $A$ and $B$. As a result we obtain  
 Eq.(\ref{pme}) immediately by noting that any separable $M\ge 0$ is a convex combination of pure product states. 
 
Now we suppose $d=pq$ is odd so that  $p\ge 3$.  If the inequality in Eq.(\ref{pmo}) were not true then from Eq.(\ref{pme}) it would follow $\tr M_{U_0}(P_q\otimes H_{L_I})=0$  for  $U_0=I_q\otimes L_I$, which would lead to, as will be shown below, the existence of a non-zero operator $R\ge0$ of the qudit with $p$ levels such that 
\begin{itemize}
\item[i.] $R$ is of rank at most two;
\item[ii.] $\tr (RV)=0$ for all $V\in \mL_p^0$ with $V\not=I$. 
\end{itemize}
In fact, since $M_{U_0}$ is separable,  we have  $M_{U_0}=\sum_{j}z_j\otimes w_j$ for some  pure states 
 $|z_j\rangle_A=\sum_n|n\rangle\otimes|x_{j,n}\rangle$ and $|w_j\rangle_B=\sum_n|n\rangle\otimes|y_{j,n}\rangle$. From $\tr M_{U_0}(P_q\otimes H_{L_I})=0$ it would follow that both Eq.(\ref{h1}) and Eq.(\ref{h2}) become now equalities, meaning that $|x_{j,n}\rangle$ should live in the subspace spanned by orthogonal states $\{|y_{j,n}\rangle, L_I|y_{j,n}^*\rangle\}$, i.e.,
$|x_{j,n}\rangle=c_{j,n}|y_{j,n}\rangle+e_{j,n}L_{I}|y_{j,n}^*\rangle$
 with $c_{j,n},e_{j,n}$ being some complex numbers,  for arbitrary $j,n$. From the distinguishability conditions  $\tr (M_{U_0}\psi_{U})=\delta_{UU_0}$  for all $U=Z_q^l\otimes L_V\in\Xi_{2d}$ we obtain
\begin{equation}\label{dcd}
\sum_j\left|{\sum}_{n\in\bZ_q}\omega^{ln}_qe_{j,n}^*\tr (L_{I}^\dagger L_{V}y^T_{j,n}) \right|^2=\delta_{VI}\delta_{l0},
\end{equation}
recalling that $y_{j,n}^T=|y_{j,n}^*\rangle\langle y_{j,n}^*|$.
From the condition Eq.(\ref{dcd}) in the case of $V=I$, it follows that $e_{j,n}^*\tr y_{j,n}^T$ is independent of $n$ and is nonzero for at least one $j$. For such a $j$, from the condition Eq.(\ref{dcd}) in the case of $V\not= I$ we obtain, taking into account  $e_{j,n}\not=0$,
$$0=\tr(L_{I}^\dagger L_{V}y^T_{j,n})=\tr (R_0+R_1^T)V$$
for all $n\in\bZ_q$ and $V\in\mL^0_{p}$, where $R_\mu=\tr_2(|\mu\rangle\langle\mu|\otimes I_p)y_{j,n}^T$ for $\mu=0,1$ with the trace taken over the qubit. We note that both $R_{0,1}\ge0$ are  of rank-1 so that  $R=R_0+R_1^T\ge0$ is at most of rank-2. 
 
 However, every nonzero $R\ge0$ satisfying $\tr (RV)=0$ for all $V\in\mL_p^0$ with $V\not=I$ is inevitably of rank 3 or more. In fact that is why we choose $\mL_p^0$. To see this we denote $R_{ab}=\langle a|R|b\rangle$ and from $\tr (RZ^a_p)=0$ for all nonzero $a\in\bZ_p$ it follows that $R_{aa}=r=\tr R/p>0$ is independent of $a$ and from $\tr (RX_pZ^a_p)=0$ for $a\in\bZ_p$ we obtain  $R_{a,a+1}=0$ for $a\in\bZ_p$. If $R$ were of a rank  at most 2, then the determinants all $3\times3$ submatrices of $R$, especially those on the diagonal with entries labeled by $\{a,a+1,a+2\}$ and $\{a,a+2,a+3\}$, would vanish so that we would have $|R_{a+2,a}|^2=r$ and $R_{a,a+3}=0$, respectively, for $a\in\bZ_p$. In the same manner, by induction, we would also have $|R_{a,a-2j}|^2=r$ and $R_{a,a+2j+1}=0$ for arbitrary $a,j\in\bZ_p$ by considering the $3\times 3$ submatrices labeled with $\{a,a+j,a+j+1\}$ and $\{a,a+j+1,a+j+2\}$. Since $p$ is odd  the equation $-2j=2j+1$ has a solution $j_p=(p^2-1)/4$ in $\bZ_p$ so that we would obtain $r=0$, i.e., $R=0$, a  contradiction  showing that every nonzero $R\ge0 $ satisfying condition ii is of rank 3 or more and thus $\tr M_{U_0}(P_q\otimes H_{L_I})>0$.


\begin{thebibliography}{99}
\bibitem{note} In \cite{lwfz14} the construction for $d=5$ refers to the numerical result in \cite{cos13}, which is confirmed here, while in the construction for $d=11$ some of those 12 states, namely $V_{3+i}$ and $V_{9+i}$,  are not orthogonal.
\end{thebibliography}
\end{document}